\documentclass{ieeeaccess}

\usepackage[font=small,labelfont=bf]{caption}
\DeclareCaptionFont{ieeeblue}{\color{accessblue}}
\DeclareCaptionLabelFormat{myformat}{\figcapfont{\textbf{#1} \textbf{#2}}}
\usepackage[flushleft]{threeparttable}
\usepackage{cite}
\usepackage{amsmath}
\usepackage{amsfonts}
\usepackage{amssymb}
\usepackage{graphicx}
\usepackage{nohyperref}
\usepackage{url} 
\usepackage{textcomp}
\usepackage[ruled,longend]{algorithm2e}

\newlength{\xfigwd}
\def\BibTeX{{\rm B\kern-.05em{\sc i\kern-.025em b}\kern-.08em
    T\kern-.1667em\lower.7ex\hbox{E}\kern-.125emX}}
\begin{document}
\history{Date of publication xxxx 00, 0000, date of current version xxxx 00, 0000.}
\doi{10.1109/ACCESS.2017.DOI}

\title{Wi-Fi and Bluetooth Contact Tracing Without User Intervention}

\author{ \uppercase{Brosnan Yuen}\authorrefmark{1}, YIFENG BIE\authorrefmark{1}, DUNCAN CAIRNS\authorrefmark{1}, GEOFFREY HARPER\authorrefmark{1}, JASON XU\authorrefmark{1}, CHARLES CHANG\authorrefmark{1}, \uppercase{Xiaodai Dong}\authorrefmark{1} \IEEEmembership{Senior Member, IEEE}, and  \uppercase{Tao Lu}\authorrefmark{1}\IEEEmembership{Member, IEEE}}
\address[1]{Department of Electrical and Computer Engineering, University of Victoria, Victoria, BC, Canada (Emails: \{brosnany, xdong, taolu\}@uvic.ca)}

\tfootnote{This work was supported in part by the Nature Science and Engineering Research Council of Canada (NSERC) Discovery (Grant No. RGPIN-2020-05938),  and Threat Reduction Agency (DTRA) Thrust Area 7, Topic G18 (Grant No.GRANT12500317), NSERC Grant 520198, Fortinet Research under Contract 05484 and NVidia under GPU Grant program.}

\markboth
{B. Yuen \headeretal: Wi-Fi and Bluetooth Contact Tracing Without User Intervention}
{B. Yuen \headeretal: Wi-Fi and Bluetooth Contact Tracing Without User Intervention}

\corresp{Corresponding author: Xiaodai Dong (e-mail: xdong@ece.uvic.ca) and Tao Lu (e-mail: taolu@ece.uvic.ca).}

\begin{abstract}

Previous contact tracing systems required the users to perform many manual actions, such as installing smartphone applications, joining wireless networks, or carrying custom user devices. This increases the barrier to entry and lowers the user adoption rate. As a result, the contact tracing effectiveness is reduced. Unlike the systems above, we propose a new privacy preserving Wi-Fi and Bluetooth (BLE) contact tracing system that does not require smartphone applications, joining wireless networks, or custom user devices. Our specially built routers seamlessly track smartphones, laptops, smartwatches, BLE headphones, and tablets without any user action, but do not trace user identity. Mapping between devices and users is only carried out for confirmed cases and suspected contacts. Moreover, we can track the absolute positions of user devices within 1.0 m due to using bidirectional long short-term memory neural networks that are trained with data pre-collected by an autonomous robot.  This allows public health authorities to track indirect droplet and surface transmissions that other contact tracing systems often overlook.

\end{abstract}

\begin{keywords}
Contact tracing, Received signal strength indicator (RSSI), Round trip time (RTT), Fine time measurement (FTM), Wi-Fi indoor localization, Bluetooth indoor localization
\end{keywords}

\titlepgskip=-25pt

\maketitle

\section{Introduction}

When a new outbreak appears with unknown pathogens, vaccines and treatments are not available immediately to reduce the spread of the disease.  Therefore, governments and public health agencies use  extensive  disease testing to identify infected individuals. However, testing the entire population is inefficient because of the limited testing capacity, false negative cases, and the associated costs. Contact tracing has  been developed to  make efficient use of the limited testing resources, where the closest contacts of the confirmed cases or symptomatic cases are tested and isolated.

\subsection{Contact Tracing}

Contact tracing is difficult because super-spreaders could infect thousands of people a day~\cite{majra2020sars} and exponentially increase the number of people in the contact tracing list. Traditionally, contact tracing has been done by hand, where the authorities interview each confirmed case to get the contacts and  visited places. Afterwards, suspected cases are isolated and tested. Symptomatic cases and high exposure cases get a higher priority in testing. With a high enough contact tracing efficiency,  diseases can be locally contained and sometimes be eradicated~\cite{eames2003contact}. However, performing contact tracing manually is very inefficient because the infected people might forget who they met and where they visited. Staff shortages, incorrect training, and slow turnaround times can also cause inefficient contact tracing.

Many countries have moved to automated means of contact tracing~\cite{kleinman2020digital,altmann2020acceptability,hinch2020effective} via smartphones, cameras, custom tracking devices, or genome sequencing.  Cameras can be used in-conjunction with facial recognition software to track individual people. Researchers collected a database of faces and applied a convolutional neural network (CNN) to classify the presences of the people in the database~\cite{nanthini2020deep}. They are able to perform contact tracing via a web interface.  Instead of only classifying faces, other researchers have used multiple cameras to track movements in real time~\cite{yaghi2020real}. Furthermore, they can determine the actual paths of the confirmed cases for contact tracing.

Genome sequencing enables contact tracing without interviewing patients or requiring tracking devices. This particularly useful for incapacitated or unidentified patients. Jennifer L. Gardy et al.~\cite{gardy2011whole} applied hierarchical clustering to sequenced genomes in order to create a genome tree of a tuberculosis outbreak. Moreover, the genome tree perfectly matches the contact traced social network created from patient questionnaires. The main disadvantage of genome sequencing is the genome tree can only be created after the patients are infected.

On the other hand, smartphones are readily available and can be used for tracking the movements of individuals. Thus,  many governments, public health agencies, and software companies have implemented smartphone applications for contact tracing.  The Singaporean government released one of the first contact tracing applications for COVID-19~\cite{doi:10.1215/18752160-8698301}. Each smartphone application broadcasts Bluetooth Low Energy (BLE) exposure notification packets containing temporary IDs of the users. Furthermore, each smartphone receives exposure notifications from all other smartphones and checks the received temporary IDs against a database of confirmed cases. If the temporary ID is in the database of confirmed cases, then the application warns the user about an exposure. Similarly, Apple and Google have developed their own contact tracing system using BLE \cite{sharon2021blind,ahmed2022privacy}, where they built contact tracing functionality into iOS and Android operating systems. This allows them  to do contact tracing on a scale of multiple countries, which is far greater than any other research study.  On the other hand, researchers have developed DigitalPPE \cite{woodward2020digitalppe}, a wearable BLE smartwatch,  that tracks social interacts between people.  DigitalPPE gives a vibration warning if two people get too close and records the IDs of the smartwatches with the relative distance.
T. Shelby et al. \cite{shelby2021pilot} performed two BLE contact tracing studies: one study using a smartphone application and the other study using external BLE tags on the user.  The custom BLE tags  had a higher accuracy compared to the smartphone application because the BLE tags had a higher transmit rate and power. Also, other researchers have relentlessly applied BLE for contact tracing~\cite{ reichert2021survey,zhao2020accuracy,leith2020coronavirus,di2021bluetooth,hernandez2020evaluating,madoery2021feature,cunche2020using,Gorji2020}.

Alternatively, T. Yasaka et al.~\cite{yasaka2020peer} used QR codes for tracking social gatherings between groups of people.  The host of the social gathering creates a QR code using the application, and the participants scan the QR code to build a time series graph.  When a user indicates a positive test result, all users within 3 traversals of the time series graph are notified.  A few more research papers have used the QR code approach~\cite{nakamoto2020qr,hoffman2020towards,mobo2020using}. Moreover, the smartphones' GPS can be used to track users in the outdoor environments~\cite{wang2020new}. This would provide a higher position accuracy than BLE and QR codes.

Wi-Fi can be a useful tool for localization and contact tracing.  A. Trivedi et al.~\cite{trivedi2021wifitrace} developed a Wi-Fi based contact tracing system without the need to install an application onto the smartphone.  They used the access points (APs) of two universities to collect packets from smartphones, where the user's trajectory is built using the closest APs. Furthermore, a graph search algorithm takes the user's trajectory and produces a location and proximity report of the exposed users.  Other research groups have used Wi-Fi based smartphone applications~\cite{li2021vcontact} to capture beacon frames from nearby APs and upload the data to the cloud. This allows the authorities to track the visited places and the positions of the confirmed cases. Moreover, the lifespan of the disease can be known  due to the recorded timestamps of the beacon frames.

\subsection{Indoor localization}

Localization is fundamental to contact tracing, and it has two major categories: outdoor and indoor localization. Out of all the outdoor localization methods, Global Positioning System (GPS) is the most popular and is robust against signal interference and jamming~\cite{6085582}. However, GPS requires direct line-of-sight (LoS) between the satellites and the handset, which is unsuitable for indoor localization.

Indoor localization has drawn more attention in the industry for its wide variety of use cases, such as autonomous indoor vehicles (AIVs)~\cite{gokhale2021feel}, unmanned aerial vehicles (UAVs)~\cite{tiemann2017scalable}, home automation, and smart buildings~\cite{moreno2016low}.  Radio Frequency (RF) waves penetrate materials like tables and walls, making RF-based indoor localization the most adopted solution.
Moreover, RF performs better than other methods~\cite{obeidat2021review}. RF-based systems employ mobile phones for capturing wireless parameters such as angle of arrival (AOA), time of arrival (TOA), and received signal strength indication (RSSI). There are two types of RF based localization methods: ranging and trilateration/triangulation, and fingerprinting. The first method requires deploying known anchor nodes with coordinate information and synchronization among nodes, while the second method does not. In this paper, we use the wireless fingerprinting approach where the RF parameters act as fingerprints for positioning. With the help of machine learning, the average localization error of fingerprinting is around 1 m~\cite{hoang2019recurrent,hoang2018soft}.

Our interests lie in estimating the positions of people to determine COVID-19 exposures. As a result, indoor localization is more useful than outdoor localization due to indoor environments having a higher infection rate ~\cite{kane2021safer}.   Moreover, indoor localization is extremely helpful for tracking COVID-19 outbreaks in complex environments such as supermarkets and airports.

\subsection{Features of the  Proposed Contact Tracing System}

\begin{table*}[ht!]
\caption{ \label{tab:featurecomparison} Feature Comparison of Indoor Contact Tracing Systems }
\begin{center}
\begin{threeparttable}
    \begin{tabular}{|p {2.9 cm}|p {1.4 cm}|p {1.7 cm}|p {1.6 cm}|p {1.6 cm}|p {1.4 cm}|p {1.9 cm}|p {1.3 cm}|}
\hline
\hline
Contact Tracing System & Tracking Method & Requires Smartphone Applications ? &  Requires Wireless Network Connections ? & Requires External User Devices ? &  Tracks Droplet and Surface Exposures ? & Accuracy & Cost \\ \hline

N. Nanthini et al.~\cite{nanthini2020deep} & Camera & No & No & No & Yes & Absolute  Position Within 0.5 m & 1 Camera per Room \\ \hline

Jennifer L. Gardy et al.~\cite{gardy2011whole} & Genome Sequencing & No & No & No & Yes & 1\% Error Rate & >\$400 USD per Sequence  \\ \hline

TraceTogether~\cite{doi:10.1215/18752160-8698301} & BLE & Yes & Yes & No & No & Relative  Position Within 1.0 m & Free \\ \hline

Apple and Google Exposure Notification \cite{sharon2021blind,ahmed2022privacy} & BLE & No, Requires Manual Activation & Yes & No & No & Relative  Position Within 1.0 m & Free \\ \hline

DigitalPPE \cite{woodward2020digitalppe} & BLE & No & No & Yes & No & Relative  Position Within 1.0 m & 1 Wearable per Person \\ \hline

T. Yasaka et al.~\cite{yasaka2020peer} & QR codes & Yes & Yes & No & No & Requires Users to Scan QR Codes  & Free \\ \hline

A. Trivedi et al.~\cite{trivedi2021wifitrace} & Wi-Fi & No & Yes & No & No & Position Within the Room & >\$340 USD per Router \\ \hline

vContact~\cite{li2021vcontact} & Wi-Fi & Optional app automates contact tracing* & Optional app automates contact tracing*  & No & Yes & Absolute  Position Within 2.0 m & Free* \\ \hline

   \textbf{ Proposed Contact Tracing System} & \textbf{ Wi-Fi and BLE} & \textbf{ No } & \textbf{ No } & \textbf{ No } & \textbf{ Yes } & \textbf{Absolute Position Within 1.0 m } & \textbf{ \$30 USD per Router} \\ \hline
    \end{tabular}
\begin{tablenotes}
\item[*] Note: If the user does not install the optional smartphone application, then a medical personnel manually performs contact tracing by revisiting every location the user has been to.
\end{tablenotes}
\end{threeparttable}
\end{center}
\end{table*}

Every contact tracing system has its own unique features and advantages, as shown in Table \ref{tab:featurecomparison}. Camera contact tracing systems \cite{nanthini2020deep} do not require any smartphone applications, wireless network connections, and external user devices. Moreover, they have an accuracy of 0.5 m and can track droplet and surface transmissions. However, setting up multiple cameras per building and a video processing system is extremely costly. Similar to the camera contact tracing system, genome sequencing \cite{gardy2011whole} is highly accurate and precise. However, it requires viral samples from each user and processing each sample is expensive.

BLE contact tracing systems are cheap and easy to set up. However, they usually require the user to manually install smartphone applications \cite{doi:10.1215/18752160-8698301} or manually activate exposure notifications in the settings \cite{sharon2021blind,ahmed2022privacy}. This results in a low participation rate and decreases the accuracy of contact tracing. Furthermore, those systems only record the relative positions of the users, of which are highly ineffective in tracking droplet and surface transmissions. Using custom BLE smartwatches \cite{woodward2020digitalppe} or tags \cite{shelby2021pilot} eliminates the need for smartphone applications, but they only record relative positions and have the exact same problems.

A few Wi-Fi contact tracing systems \cite{trivedi2021wifitrace} do not require the user to install smartphone applications. Instead, the routers record the positions of the smartphones, whenever the user manually logs into the wireless network. This approach has low position accuracy due to users getting disconnected or logging out. Furthermore, the RSSI ranking system has location ambiguity due to multiple positions having the same RSSI ranking. As a result, they can not determine if a user is within 2.0 m of another user. Moreover, their system is expensive due to them using Cisco and HP/Aruba equipment that cost >\$340 USD per router.

Unlike the previous systems in Table \ref{tab:featurecomparison}, we propose a new privacy preserving Wi-Fi and BLE indoor contact tracing system that does not require the users to perform any actions. Specifically, the users do not need to install any smartphone applications. The users do not need to connect to any wireless networks, which improves localization accuracy due to eliminating wireless network disconnects and users logging out.  Instead, we use custom designed ESP32C3 routers to capture Wi-Fi and BLE packets emitted from the wireless devices. The overall system contains four modules: an autonomous robot for site survey, a BiLSTM network for trajectory prediction, WiFi routers designed with special features to collect sufficient RF data and pre-process the data, and graph based contact tracing algorithm. In each module, there are innovations in design solution and practical implementation, as detailed in the next sections. This system allows us to track smartphones, smartwatches, tablets, and laptops of users seamlessly in the background. Although user tracking and device tracking are used interchangeably throughout the paper, the system does not obtain user identity for privacy purpose but only trace device WiFi interface MAC addresses. Our contact tracing system also tracks droplet and surface transmissions due to our neural networks providing absolute positions. Subsequently, indirect or delayed infections can be tracked even-though the infected individual has left the area multiple days ago.  As for localization accuracy, our contact tracing system has an average error of 1.0 m, which is similar to the other BLE and Wi-Fi contact tracing systems. However, the camera and genome sequencing methods have higher localization accuracy at a cost of much more expensive equipment.

The paper is organized as follows. Section~\ref{proposed} is a big picture overview of the proposed contact tracing system. The site survey is conducted in Section~\ref{sitesurvey}, while the data processing is shown in Section~\ref{packetprocessing}. The actual contact tracing algorithm is depicted in Section~\ref{contacttracingsec}. Section~\ref{results} shows the results and discussions of identifying unique mobile devices, localization performance, and contact tracing. A conclusion is presented in Section~\ref{con}.

\section{Overview of the Proposed Contact Tracing System}\label{proposed}

\begin{figure}[!t]
\centering
\includegraphics[width=0.5\textwidth]{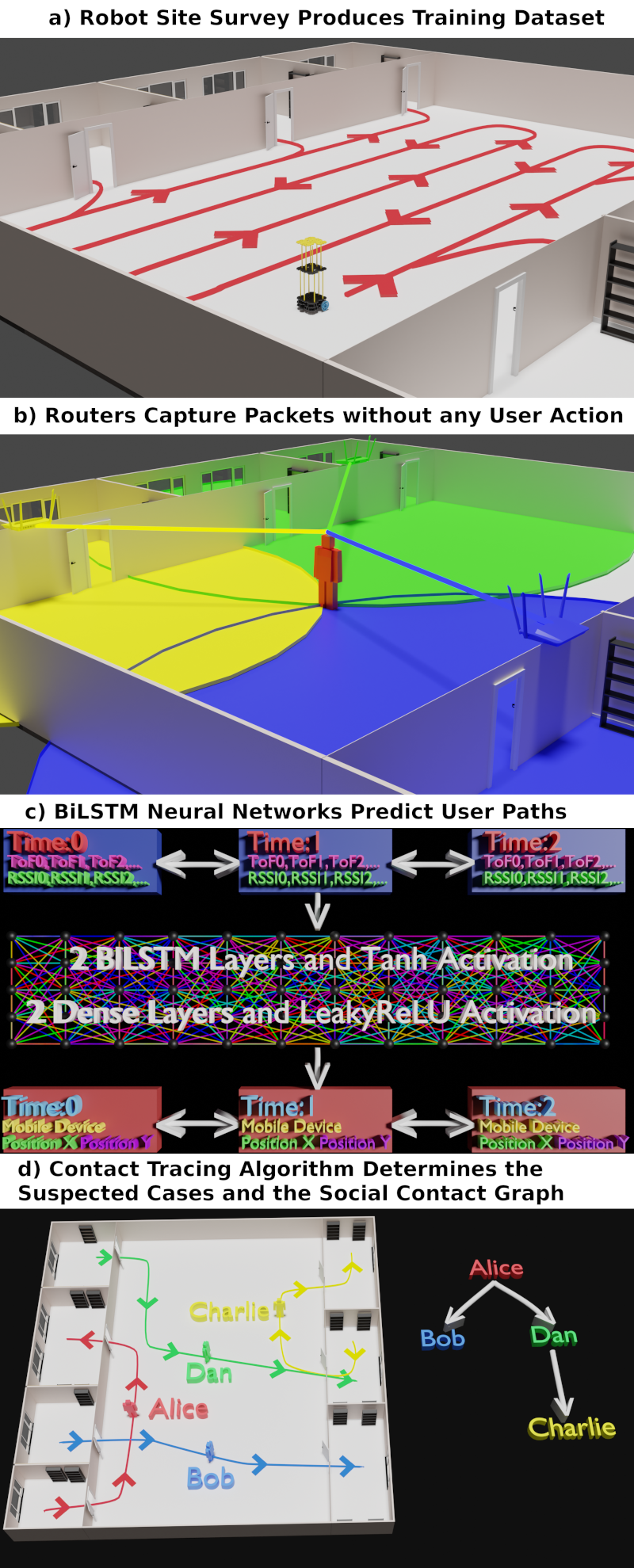}
\caption{Overview of the proposed contact tracing system.} 
\label{fig:sysdiagram}
\end{figure}

 Fig.~\ref{fig:sysdiagram} shows the overview of the proposed contact tracing system. It consists of four components: 1) An autonomous robot for site survey to generate a location-fingerprint database; 2) A BiLSTM neural network trained by the site survey dataset for user trajectory prediction; 3) WiFi routers for capturing packets without user action in prediction, testing and training stages; 4) Contact tracing algorithm and engine based on the localization data.  
 
 We propose to use a bidirectional long short term memory (BiLSTM) neural networks to predict the trajectories of mobile devices. The BiLSTM requires many datasets such as the training dataset, the testing dataset, and the production dataset. The training dataset is used to train the BiLSTM, while the testing dataset measures the accuracy of the BiLSTM. Moreover, the production dataset is the real life dataset that only contains  input features without any labels. 
 
 Wireless fingerprinting for localization does not need to install known anchor nodes but does need to have a location-fingerprint database of a site. This site survey if done manually is very laborious. The purpose of the Turtlebot3 site survey is to obtain the training dataset and the testing dataset using a robot. The Turtlebot3 executes autonomous site surveys by meticulously visiting all positions on the floor. A smartphone is mounted on the robot, and it broadcasts wireless packets while moving in order to simulate mobile device trajectories. On the other hand, the ESP32C3 routers capture Wi-Fi and BLE packets for the training dataset, the testing dataset, and the production dataset. For the production dataset, the Turtlebot3 is not involved, and the routers directly capture packets from the users  without the users needing to perform any action. Afterwards, the packets' transmit power (TX power), received signal strength indication (RSSI), and time of flight (ToF)  are used to predict the user trajectories.

Finally, we design a graph based contact tracing algorithm to build a social contact graph. Every user is assigned to a unique node on the graph. For every intersection between the trajectory of a confirmed case and the trajectory of a user, we add an edge that connects the node of the confirmed case to the node of the user. After repeating this process  multiple times, a graph of the suspected cases is displayed together with their trajectories.

In the next sections, each component of the system is described in details.

\section{Turtlebot3 for Site Survey}\label{sitesurvey}

Typically, mobile devices transmit many Wi-Fi and BLE packets as they move around the building. By implementing packet sniffing on the router side, mobile devices can be tracked throughout the day. However, the localization algorithms require large amounts of training and testing data. Measurement of the training data is done in the form of a site survey, where a mobile device transmits packets at every position and the signal information is recorded at the router side.

Collecting data by hand is extremely tedious and introduces position errors. Instead, we built a custom Turtlebot3~\cite{amsters2019turtlebot} for the site survey. The Turtlebot3 continuously transmits packets to the routers, while visiting every position in the building. The original Turtlebot3 has a height of 19 cm, which is too short for the height of a smartphone on a table or in a user's pocket. A platform is added to the custom Turtlebot3 in order to increase the smartphone's height to 75 cm. Moreover, the custom Turtlebot3 is also equipped with RPLIDAR A2, Intel D415 RGBD camera, Nvidia Jetson TX2, and Raspberry Pi 3.

\subsection{Robot Operating System 2}

Robot Operating System 2 (ROS2)~\cite{erHos2019ros2} is an open-source robotics framework that collects sensor information,  executes data processing, implements inter-process communications, and allows real-time control. ROS2 has four main concepts: nodes, topics, services, and actions. Nodes are individual processes, of which execute a singular task like collecting sensor data or filtering information. Nodes can commence one way communications with other nodes by publishing messages to topics. All nodes that subscribe to a specific topic receive the same messages. Unlike topics, services are a two-way communications channel. Nodes can send service requests and receive   service responses once the specific operation is completed. Actions are an extension of services, where the nodes receive periodic feedback status messages instead of not receiving feedback messages.

The RPLIDAR A2 is a 2D laser ranging device that measures the distances to the nearest opaque objects. It is a 360° LIDAR that completes 1 revolution every 0.1 seconds. The 360° scans are divided into 360 angle intervals. For each angle interval, the RPLIDAR A2 returns a distance value. The laser scans feed into  ROS2 SLAM\_toolbox, of which it produces a 2D grid map and  publishes the transform from map to odometry (odom). It essentially determines the position and orientation of the Turtlebot3. On the other hand, the Intel D415 RGBD is used for obstacle avoidance.  The D415 produces a RGBD point cloud at 720p 30 frames per second (FPS). Camera sensors have false positive readings,  where the sensor outputs a ghost point in the absence of objects. In order to eliminate the false positives, multiple RGBD point cloud frames are joined together and are uniformly decimated. Afterwards, the point cloud is organized into clusters, where the cluster centres and standard deviations are calculated. If a point is 1 standard deviation away from the cluster centre, then it is removed.

\subsection{Detecting Obstacles in the Local Map}

In order for the routers to collect data, the Turtlebot3 needs to visit every position on the floor and stand there for a few minutes. To accomplish that, we create an autonomous navigation algorithm  that is able to avoid static objects and moving obstacles in order to operate in various buildings and environments. There are 4 main steps in Algorithm~\ref{alg:autonomousnav} : detecting obstacles in the local map, finding the start positions of the paths, generating the best path using the start positions, and moving along the planned path.

\begin{algorithm}
  \caption{Path Planning and Autonomous Navigation}\label{alg:autonomousnav}
  \SetKwFunction{FMain}{constructCostmap}
  \SetKwProg{Fn}{function}{:}{}
  \Fn{\FMain{}}{
  
  Odom $\leftarrow$ getOdom();
 
  LaserScans $\leftarrow$ getLIDAR();
  
  PointCloud $\leftarrow$ getCamera();
  
  Costmap $\leftarrow$ newCostmap(Costmap,LaserScans);
  
  Costmap $\leftarrow$ newCostmap(Costmap,PointCloud);
  
  }
  \Return{Costmap, Odom}

  \SetKwFunction{FMain}{detectObstacles}
  \SetKwProg{Fn}{function}{:}{}
  \Fn{\FMain{Costmap, Position}}{
  
  \For{\textbf{each} OccupiedPosition \textbf{in} Costmap}{
    Distance $\leftarrow$ magnitude(OccupiedPosition - Position);
    
    \If{Distance < RobotRadius}{
        \Return{HasObstacles}
    }
  }
  }
  \Return{NoObstacles}

  \SetKwFunction{FMain}{findStartPos}
  \SetKwProg{Fn}{function}{:}{}
  \Fn{\FMain{Costmap, Odom}}{
  
  Rays $\leftarrow$  Costmap - Odom;
  
  StartPositionList $\leftarrow$ [];
  
        \For{\textbf{each} RayVec \textbf{in} Rays}{
    Detect $\leftarrow$ detectObstacles(Costmap,RayVec);
    
    \If{Detect == NoObstacles}{
    
    RandNum $\leftarrow$ randomUniform(0, 1);
    
      Position $\leftarrow$ RandNum*RayVec;
      
      StartPositionList.append(Position);
    }
  }
  }
  \Return{StartPositionList}

  \SetKwFunction{FMain}{createPath}
  \SetKwProg{Fn}{function}{:}{}
  \Fn{\FMain{StartPositionList, Costmap, VisitedPositionList}}{

  BestPath $\leftarrow$ initPath();
  
  BestPathScore $\leftarrow$ 0;
  
  \For{\textbf{each} StartPos \textbf{in} StartPositionList}{
  
      \For{\textbf{each} EndPos \textbf{in} Costmap}{
        Direction $\leftarrow$ EndPos - StartPos;
        
        Num $\leftarrow$ Int(magnitude(Direction)/0.5);
        
        Path $\leftarrow$ arange(0,1,Num)*Direction + StartPos;

        Detect $\leftarrow$ detectObstacles(Costmap,Path);
        
        \If{Detect == HasObstacles}{
          continue;
        }
        PathScore $\leftarrow$ computeScore(Path,VisitedPositionList);
        
        \If{PathScore 
        $>$ BestPathScore}{
          BestPathScore $\leftarrow$ PathScore;
          
          BestPath $\leftarrow$ Path;
        }
      }
  }

  }
  \Return{BestPath}

  \end{algorithm}

  \begin{algorithm}

  \SetKwFunction{FMain}{moveAlongPath}
  \SetKwProg{Fn}{function}{:}{}
  \Fn{\FMain{BestPath, Odom, Costmap, VisitedPositionList}}{

  \For{\textbf{each} Position \textbf{in} BestPath}{
      Detect $\leftarrow$ detectObstacles(Costmap,Position);
      
      \If{Detect == HasObstacles}{
          moveRobotToPosition(Odom);
          
          break;
    }
      
      moveRobotToPosition(Position);
      
      sleep();
      
      VisitedPositionList.append(Position);
  }
  
  }
  \Return{VisitedPositionList}
  
\end{algorithm}

Before planing a path and moving along it, the Turtlebot3 needs to determine  the obstacles in the local vicinity. The function constructCostmap retrieves the odometry, 2D laser scans (LaserScans), and 3D point cloud (PointCloud) to construct a 2D occupancy grid (Costmap). Each pixel in the occupancy grid has a status of occupied, free space, or unknown depending on the sensor information. This helps the robot to avoid obstacles, which are labelled as occupied.

Another function, detectObstacles determines if an input position is near any obstacles in the costmap. It checks every occupied position in the costmap to determine if any of them are closer to the input position than the robot radius. If there is an obstacle closer than the robot radius, then the function indicates the presence of an obstacle by returning HasObstacles, otherwise there are no obstacles and the function returns  NoObstacles.

\subsection{Finding the Start Positions of the Paths}

Prior to generating new paths, we must first select the start positions of the paths. As shown in the function  findStartPos, rays are created starting at the Turtlebot3's current position (Odom) and ending at the occupied positions in the costmap. For each individual ray, it is checked against the costmap for potential obstacles.  If the ray has an obstacle in its path, then the ray is discarded. Afterwards, a random position is selected from the ray's line and is appended to the list of start positions (StartPositionList). The list of start positions represent candidate positions for generating the best path.

\subsection{Generating the Best Path using the Start Positions}

After finding the start positions, we use the function createPath to  generate new paths and score them for finding the best path. One of the inputs to the function is a list of previously visited positions (VisitedPositionList), of which is used to avoid visiting the same locations. Subsequently, the best path (BestPath) and best path score (BestPathScore) are initialized with the worst possible path. A low path score indicates an undesirable path, while a high path score indicates a desirable path. Afterwards,  the algorithm generates every possible combination of candidate path starting from the start position list and ending at an obstacle in the costmap.  The candidate path is a straight line that has evenly spaced points between the start position and end position. Each position in the candidate path has a minimum distance of 0.5 m to the other positions.  If a candidate path has an obstacle in its way, then it is rejected. A path score is computed for each candidate path, depending on the length of the path and the number of overlaps between the candidate path and the previously visited positions.  When the candidate path goes through many of the previously visited positions, its path score is decreased.  However, if the candidate path goes to  unvisited locations, then its path score is increased. Longer paths also increase the path score. The candidate path with the highest path score is selected as the best path.

\subsection{Moving Along the Planned Path}

As depicted in the function moveAlongPath,  ROS2 Navigation 2 is used to move the Turtlebot3 along the best path. Navigation 2 controls the velocity and angular velocity in order to visit all the evenly spaced points in the best path. Before moving to a new location, the algorithm checks to see if there is an obstacle in the way. Upon detecting a blockage in its path, Turtlebot3 retraces its steps back to the start position (Odom). When the Turtlebot3 arrives at one of the planned positions, the current position is appended to the list of  previously visited positions (VisitedPositionList). Subsequently, the robot waits at that location for a few minutes, while the ESP32C3 routers collect Wi-Fi and BLE packets from the Turtlebot3. Moreover, the ESP32C3 routers record the Unix times and positions of the robot for the training and testing databases. The process above repeats until all free positions on the map are sampled by the Turtlebot3.

\section{Packet Processing}\label{packetprocessing}

\subsection{ESP32C3 Router}

In order to capture packets, we built a custom router based on the ESP32C3 chip-set because it supports monitor mode on Wi-Fi and BLE simultaneously. It is able to determine RSSI from all Wi-Fi/BLE packets and supports Wi-Fi FTM to get the round trip time (RTT). For some BLE packets, the ESP32C3 provides the TX power of the mobile devices. An RF front end is added to increase the RX power and the dynamic range of the received packets, while a SD card is added to store the data.

 At the start of the day, the routers' real-time clocks (RTCs) are synchronized via simple network time protocol (SNTP) to the master server. The non FTM packets are timestamped by the RTCs with an accuracy of  1~${\rm\mu}$s. However, the FTM packets have a timestamp accuracy of 1 ns provided by the ESP32C3's high-resolution timer. Subsequently, raw packet data is written live to the SD card, and it is sent back to the master server at the end of the day. Most of the packet processing is done at the master server to reduce CPU load on the ESP32C3.

\subsection{Increasing Wi-Fi Response Rate using the ESP32C3}

Manufacturers limit the power consumption of the Wi-Fi chipsets on the mobile devices to conserve battery charge. As a result, the number of packets transmitted by the mobile devices is small, and the localization accuracy is low. However, we can increase the localization accuracy by sending packets to the mobile devices using the ESP32C3 and getting a higher response rate.  In order to cover most Wi-Fi channels, the routers alternate between channels 1, 6, and 11. After switching to a channel, the ESP32C3 transmits a request-to-send (RTS) packet to a mobile device. In response, the mobile device transmits a clear-to-send (CTS) packet back to the router, which contains information pertaining to the mobile device. The ESP32C3 can also send NULL packets to get an ACK response from the mobile devices. Moreover, special mobile devices can respond to Wi-Fi FTM requests, of which greatly increases the localization accuracy.

\subsection{Increasing BLE Response Rate using the ESP32C3}

Similar to the Wi-Fi case, the ESP32C3 alternates between BLE channels 37, 38, and 39. Upon switching to a channel, the router sends a BLE scan request to a mobile device. Afterwards, the mobile device replies with a BLE scan response, of which contains BLE capabilities and sometimes model specific information. If the mobile device advertises a BLE service, then the ESP32C3 will send a pairing request packet to the device. Even if the pairing request is denied, the router will still receive a pairing response packet and the position of the mobile device.

\subsection{Sorting Wi-Fi Packets by Source Type}

Once the packets from the routers are collected, they are sorted and processed. The routers record the packets in non-chronological order. Therefore, the packet processor sorts the packets by timestamp in order to synchronize the packets from multiple routers. Note that malformed Wi-Fi packets are discarded due to having incorrect information. Afterwards, the packets are sorted by the type of wireless device: AP, wireless distribution system (WDS), bridged device, or mobile device. APs are found by looking at the source MAC addresses of the beacon frames. WDS are identified when the packets have ToDS=1 and FromDS=1. Bridged devices are identified when the packets have FromDS=1 and the source MAC addresses do not equal the BSSIDs. The remaining wireless devices are categorized as mobile devices. For the purposes of this paper, only the packets from the mobile devices are used for contact tracing.

\subsection{Sorting BLE Packets by Source Type}

Same as the Wi-Fi packet sorting and processing, the BLE packets are sorted by their arrival time. Subsequently,  BLE packets with invalid cyclic redundancy check are discarded due to having incorrect protocol data unit types and incorrect manufacturer specific information. Afterwards, BLE packets are sorted by the type of TX address: public MAC addresses and random MAC addresses.  Public BLE MAC address are stable and constant for long periods of time, so they are easily tracked and localized. On the other hand, random BLE MAC addresses rapidly change from one packet to another packet, and they require special BLE MAC de-randomization algorithms for tracking.

\subsection{Defeating Wi-Fi MAC Address Randomization}

Many mobile devices randomize their Wi-Fi MAC addresses to prevent user tracking and identification~\cite{martin2017study}.  In order to defeat Wi-Fi MAC address randomization, we create an algorithm to categorize mobile devices using model specific information from the probe requests. Firstly, we capture probe request frames emitted by the mobile devices. Every mobile device regularly transmits probe request packets, so this is not a problem. Secondly, we extract model specific information from the probe requests.  Each device model type has unique model specific information such as supported rates, extended supported rates, high throughput (HT) capabilities, direct sequence (DS) parameter set, and vendor specific organizationally unique identifier  (OUI). Furthermore, those model specific information are fixed and do not change over the lifetime of the device~\cite{robyns2017macfingerprinting}.

Thirdly, the model specific information is converted into a binary fingerprint vector. For example, if transmit beamforming is supported on the device, then it is set to "1" in the binary fingerprint vector, otherwise it is set to "0". In order to make the binary fingerprint vectors the same length, missing values are padded with "0". Fourthly, we use the binary hamming distance to compare binary fingerprint vectors. Even though two packets might have completely different MAC addresses, if the hamming distance of two packets'  binary fingerprint vectors  is zero, then the two packets originated from the same model type.  This allows us to track and locate individual model types by collecting packets with the same  binary fingerprint vectors. Finally, we use a ball tree clustering algorithm  to categorize binary fingerprint vectors into their respective device types. The ball tree is constructed such that each leaf node contains the exact same binary fingerprint vector. Moreover, each branch contains device types from the same device family. As a result, we can cluster unknown and new device types around well known device types.

\subsection{Extracting Features from Wi-Fi Packets For Localization}

Specific Wi-Fi packet features are extracted as fingerprints for localization. Wi-Fi RSSI and Wi-Fi Signal Quality Index (SQI) correlate to the received  (RX) powers of the routers. High RSSI values indicate the mobile devices are close to the routers and the RX power of the routers is large. Low RSSI values indicate mobile devices are far away from the routers and the RX power of the routers is small. Some wireless interfaces report the noise power of specific channels and packets. SQI is a function of RSSI and noise power. If the RSSI values are high, then the SQI values are high. Moreover, if the noise power is high, then the SQI values are low.

 RSSI and SQI are susceptible to the type of mobile device, transmit (TX) power, and noise power. This makes RSSI and SQI somewhat unstable and dependent on the environmental conditions. On the other hand, some mobile devices support Wi-Fi FTM, of which greatly increases the robustness. To initiate FTMs, the router sends an FTM packet to the mobile device containing the FTM packet's departure time. Afterwards, the mobile device sends an ACK packet to the router containing the FTM packet's arrival time and the ACK packet's departure time. The router records the ACK packet's arrival time and computes the RTT. In order to obtain a more precise RTT, multiple series of FTM exchanges are performed and are averaged. ToF can be computed from RTT, and ToF is invariant to the type of mobile device, TX power, and noise power. In conclusion, ToF is far less susceptible to the external environment when compared to RSSI and SQI.

\subsection{Extracting Features from BLE Packets For Localization}

The main disadvantage of Wi-Fi packets is the lack of TX power information. Every mobile device has a different TX power, and it results in different RX powers.  Inconsistent RX powers produce incorrect localization predictions and  invalid contact tracing paths. This is an open problem. We propose to use BLE packets to assist the Wi-Fi RX power calibration. Note that BLE packets contain the information of both TX powers and RX powers. 
The BLE path loss $L_{BLE}$
\begin{equation}
    L_{BLE}(\Vec{X}) = P_{BLETX} - P_{BLERX}(\Vec{X})
\end{equation}
is computed using the constant BLE TX power $P_{BLETX}$ and the BLE RX power $P_{BLERX}$ as a function of position $\Vec{X}$. It is invariant to the type of mobile device, since the path loss only depends on the distance to the router and the channel environment. Thus, the BLE path loss $L_{BLE}$ is used as one of the inputs to the neural networks. Assuming the BLE path loss is equal to the 2.4 GHz Wi-Fi path loss at the same position $L_{BLE}(\Vec{X}) = L_{WiFi}(\Vec{X})$, the Wi-Fi TX power $P_{WiFiTX}$
\begin{equation}
    P_{WiFiTX} =  L_{BLE}(\Vec{X})   +   P_{WiFiRX}(\Vec{X}) 
\end{equation}
can be calculated. Since the  Wi-Fi TX power $P_{WiFiTX}$ is constant for a specific model of mobile device,  the Wi-Fi path loss $L_{WiFi}$
\begin{equation}
    L_{WiFi}(\Vec{X}) = P_{WiFiTX} - P_{WiFiRX}(\Vec{X})
\end{equation}
can be computed. As a result, localization predictions using path loss have a higher accuracy.

\subsection{Neural Networks for Localization}

After extracting the features from the packets, they are fed into the BiLSTM neural networks to predict the positions of mobile devices. However, the BiLSTM neural networks require a large number of continuous trajectories for training.  The training trajectories are generated via a simple recursive algorithm from the site survey sample locations. Firstly, a random position is selected as the current  position $\Vec{P}_i$. Secondly, another random position is selected as the candidate position $\Vec{P}_C$ for the next position $\Vec{P}_{i+1}$. If the Euclidean distance between the candidate position  and the current position is less than a distance threshold $|\Vec{P}_C - \Vec{P}_i| <  |k| $, then the candidate position becomes the next position $\Vec{P}_{i+1} \leftarrow \Vec{P}_C  $; otherwise a new random position is selected as the  candidate position $\Vec{P}_C$. The distance threshold $k$ is a random normal number that has a standard deviation of 1 m. Finally, the process above repeats until a trajectory of positions $ \{ \Vec{P}_{0}, \Vec{P}_{1} , \Vec{P}_{2}, \ldots , \Vec{P}_{N}  \}$ is completed.  For each floor in the building,  20,000 trajectories are randomly generated for training. Moreover, each trajectory has 20 different positions.

Once the training dataset is generated, it is used to train the BiLSTM neural networks. The mentioned neural networks have special neurons because they can retain  information such as the past positions and the past features. This allows the BiLSTM to predict future positions using past positions and past features. Moreover, the inverse is also true because the BiLSTM can use future positions and future features to predict past positions. If ToF or SQI features are available, then they are fed into the network as a time series of features. If TX power is present, then signal path loss is used as an input feature to the network. When the features listed above are not available, the neural network defaults to RSSI for predicting the trajectories of mobile devices.

 As shown in Fig.~\ref{fig:sysdiagram}, the BiLSTM neural network consists of 2 BiLSTM layers followed by 2 dense layers. The number of input features to the network is denoted by $F_{input}$ as it changes depending on the number of routers. Each BiLSTM layer consists of $7F_{input}$ neurons with tanh activation functions. The first dense layer has $14F_{input}$ neurons with LeakyReLU activation functions, while the second dense layer has 2 neurons with no activation function. Furthermore, the second dense layer outputs the predicted positions of a mobile device.

\section{Contact Tracing Algorithm}\label{contacttracingsec}

  \begin{algorithm}
  \caption{Contact Tracing}\label{alg:contactracing}
  
  \SetKwFunction{FMain}{lookupConfirmedCase}
  \SetKwProg{Fn}{function}{:}{}
  \Fn{\FMain{Database, InitialTime, InitialPos, MACaddr, ModelName}}{

UserInfos $\leftarrow$ getValues(data=Database, key=InitialTime);

\eIf{MACaddr != NULL}{
UserInfos $\leftarrow$ getValues(data=UserInfos, key=MACaddr); 
}{
UserInfos $\leftarrow$ getValues(data=UserInfos, key=ModelName); 
}

  MinDistList $\leftarrow$ [];
  
  \For{\textbf{each} User \textbf{in} UserInfos}{
     AbsPos $\leftarrow$ square(IntialPos - User.Pos);
     
     Dist $\leftarrow$ sum(AbsPos, 2);
     
     Index $\leftarrow$ minIndex(Dist);
     
     MinDistList.append(Dist[Index]);
  }
  
  Index $\leftarrow$ minIndex(MinDistList);
  
  }
  \Return{UserInfos[Index]}

  \SetKwFunction{FMain}{pathIntersect}
  \SetKwProg{Fn}{function}{:}{}
  \Fn{\FMain{UserInfo1, UserInfo2, DistanceThres, TimePeriodThres    }}{
  DoesIntersect $\leftarrow$ False;
  
  TimeMatrix  $\leftarrow$ transpose(UserInfo1.Times) - UserInfo2.Times;
  
  TimeMatrix  $\leftarrow$ abs(TimeMatrix);
  
  Index $\leftarrow$ select(TimeMatrix $<$ TimePeriodThres);  
  
  PosMatrix  $\leftarrow$ transpose(UserInfo1.Pos[Index]) - UserInfo2.Pos[Index];
  
  PosMatrix  $\leftarrow$ square(PosMatrix);
  
  PosMatrix  $\leftarrow$ sqrt( sum(PosMatrix, 3) );
  
  Index $\leftarrow$ select(PosMatrix $<$ DistanceThres);  
  
  \If{Index.size $>$ 0}{
DoesIntersect $\leftarrow$ True;
 }
  
  }
  \Return{DoesIntersect}

  \SetKwFunction{FMain}{findContacts}
  \SetKwProg{Fn}{function}{:}{}
  \Fn{\FMain{InputUsers, UserInfoList,  AdjMat}}{

  UserSize $\leftarrow$ UserInfoList.size;
  
  OutputUsers  $\leftarrow$ [];
  
    \For{i $\leftarrow$ 1 
 \textbf{To} UserSize}{
 
    TargetUser $\leftarrow$ UserInfoList[i];
    
    \If{!InputUsers.contains(TargetUser)}{
    continue;
    }
     
         \For{j $\leftarrow$ 1 
     \textbf{To} UserSize}{
     
         CurrentUser $\leftarrow$ UserInfoList[j];
     
         Result $\leftarrow$ pathIntersect(TargetUser, CurrentUser);
         
         AdjMat[i][j]  $\leftarrow$ Result;
         
        \If{Result}{
        OutputUsers.append(CurrentUser);
        }
         
      }
     
  }
    
  }
  \Return{AdjMat, OutputUsers}

  \end{algorithm}

  \begin{algorithm}

  \SetKwFunction{FMain}{createGraph}
  \SetKwProg{Fn}{function}{:}{}
  \Fn{\FMain{ConfirmedCases, UserInfoList, SearchDepth   }}{
  
  UserSize $\leftarrow$ UserInfoList.size;

  AdjMat $\leftarrow$ zeroMatrix(UserSize, UserSize);
  
  InputUsers $\leftarrow$ ConfirmedCases;
  
  \For{Depth $\leftarrow$ 1 
 \textbf{To} SearchDepth}{
     AdjMat, OutputUsers $\leftarrow$ findContacts(InputUsers, UserInfoList,  AdjMat);
     
      InputUsers $\leftarrow$ OutputUsers;
  }
  
   }
  \Return{AdjMat}

\end{algorithm}

The contact tracing algorithm takes a mobile device's MAC address/model name, an initial time and/or an initial position of a confirmed case as input and allows us to precisely track the paths of confirmed and suspected cases with an accuracy of 1.0 m. Furthermore, we can track droplet and surface exposures due to knowing the absolute positions of the users. The contact tracing procedure described in Algorithm \ref{alg:contactracing} consists of 4 main parts: the key-value database system, looking up the paths of the confirmed cases, finding the suspected cases using path intersection, and creating a graph connecting the confirmed cases to the suspected cases.

\subsection{Key-Value Database System}

For simplicity, a key-value database system is used to store the user information for the contact tracing system. Given a unique key, the algorithm can use it to look up a specific value in the database.  Searching for values by comparing each key individually is extremely slow because it takes $O(N)$ time.  However, hashmaps can decrease the search time to $O(1)$ constant time. Our system is built on Redis, a hash based key-value database system, where it hashes the unique key to obtain a pointer. Afterwards, the pointer is used to access the memory location of the associated value. In particular, Redis uses the cyclic redundancy check (CRC) hash function family to lookup key-value pairs because it is simple and has hardware acceleration in modern CPUs. As a result, Redis speeds up the user information look-ups in the contact tracing system.

The contact tracing database contains many key-value pairs. Each value contains the user's device model name, MAC addresses, positions, date/time, medical test results, and contacts with other users. Furthermore, an initial position or a MAC address can be used as a key to look up those values. These properties are useful for looking up the paths of the confirmed cases.

\subsection{Looking Up the Paths of the Confirmed Cases}

The first step of the contact tracing algorithm is to lookup  user information of the confirmed cases. However, many users have the exact same identifiers such as the same trajectory, the same device model name, and the same random MAC addresses. As a result, significant ambiguity is present in the lookup process. To solve the problem above, we create the lookupConfirmedCase function to reduce the identifier ambiguity. There are 4 main scenarios, where the function has to perform look-ups:

 Scenario A: all mobile devices do not have MAC address randomization.  This causes each mobile device to have a single unique MAC address that is easily identified and tracked by the algorithm. When a patient has a positive test result, they only need to provide their single MAC address (MACaddr) and initial time (InitialTime) to look-up user information (UserInfos). The initial time requires the specific day and hour of the arrival in the location.

  Scenario B: all mobile devices have MAC address randomization, but each mobile device has a unique model name.  Due to the fact that each unique model name emits a unique probe request signature, we can still identify and track individual mobile devices. This time, the patient has to provide the device model name (ModelName) and the initial time. For example, the device model name could be iPhone 13, Galaxy S22, or Pixel 6. As a result, the algorithm can look up the information on the confirmed case without knowing the actual MAC addresses.

  Scenario C: all mobile devices have MAC address randomization and multiple devices have the exact same model name. However, devices with the same model names have unique trajectories that do not have intersecting points with each other. This scenario is far more difficult than Scenario B, and the algorithm can only tell users apart by their unique trajectories. Thus, the patient needs to provide the initial position (InitialPos), model name, and initial time. The input initial position could be any position on the patient's trajectory. The algorithm selects the user information that contains the closest trajectory to the initial position.

  Scenario D: all mobile devices have MAC address randomization and multiple devices have the exact same model name. Furthermore, multiple devices with the same model names have intersecting trajectories with each other or near misses. In some situations, two physically separated trajectories might be mislabelled as having an intersecting point because the neural networks predicted the wrong positions. Subsequently, the algorithm falsely groups multiple individual users as a single confirmed case. This increases the false positive rate and adds more users to the list of suspected cases. On the other hand, the false negative rate stays the same because the algorithm still traces the correct social contacts.

\subsection{Finding the Suspected Cases using Path Intersection }

After retrieving the user information of the confirmed cases, we use the pathIntersect function to find  the suspected cases. Given the paths of User1 and User2, the function  determines if their paths intersect within a certain distance and time threshold. Firstly, we compute the time differences (TimeMatrix) between  both user paths. Secondly, we select specific positions (PosMatrix) from the user paths that have time differences less than the time period threshold (TimePeriodThres).  Typically, public health authorities will input a different time period threshold for each type of pathogen. Thirdly, we select positions from the user paths that are closer than the distance threshold (DistanceThres). We set the distance threshold equal to 2.0 m because our localization accuracy is around 1.0 m. If there exists at least one distance less than the distance threshold, then the function indicates an intersection, otherwise the function does not indicate an intersection.

\subsection{Creating a Graph Connecting the Confirmed Cases to the Suspected Cases}

Using the path intersection function, we create a graph connecting the confirmed cases to the suspected cases. A graph is defined as a set of nodes that are connected by edges. The adjacency matrix $AdjMat$ describes every edge connection, where $AdjMat[i][j]=True$ represents a connection between node $i$ and node $j$. On the other hand, $AdjMat[i][j]=False$ represents no connection between  node $i$ and node $j$. In particular,  each user is assigned to a unique node and each social contact is represented by an edge connection.  To begin, the function createGraph retrieves the total number of users (UserSize), and the adjacency matrix is initialized as a zero matrix of size UserSize by UserSize. Afterwards, the list of confirmed cases (ConfirmedCases) is selected as the list of input users (InputUsers). The  findContacts function compares the input users' paths to every other user path in the list of total users. If their paths intersect, then the corresponding element in the adjacency matrix is updated $AdjMat[i][j]=True$ and the new social contact is appended to the output list (OutputUsers). Pathogens can spread very rapidly due to infecting their primary contacts and later the secondary contacts of those primary contacts. The contact tracing algorithm gets ahead of the disease spread by recursively applying the findContacts function until the search depth (SearchDepth) is reached. This produces a social contact graph that is very deep and has many degrees of separation. In conclusion, the createGraph function returns the fully built adjacency matrix connecting the confirmed cases to the suspected cases.

\subsection{User Privacy Considerations}

Privacy is a very important aspect to keeping collected information safe and within regulations with Canadian and British Columbia Privacy Acts. In our system, phone numbers, email addresses, and legal names are not collected by the routers and are not stored in the database. Only the MAC addresses of the Wi-Fi/BLE chipsets and device model names are obtained as the identification of the mobile devices. Note that MAC addresses cannot directly identify users, and MAC address randomization also complicates the mapping of multiple MAC addresses to user devices. Public health authorities will only map the MAC addresses to user identities for confirmed cases and suspected contact cases, with the help of additional information of user identity and device wireless interface MAC addresses. 
Furthermore, as part of the privacy protection, users that enter a building  with the contact tracing system in place need to be aware of what the system does and actively consent to their data being collected. Users also have the ability to retroactively erase their information in the contact tracing database. Finally, the collected data such as MAC addresses, device model name, and  positions are encrypted with AES256 algorithm.

\section{Results and Discussion}\label{results}

Multiple datasets were collected at the University of Victoria, Victoria, British Columbia, Canada  in the Engineering Office Wing (EOW) 3rd floor, EOW 4th floor, Engineering Computer Science (ECS) 1st floor, and ECS 5th floor. At each floor, the Turtlebot3 physically moves along 3 unique trajectories. Every trajectory contains unique positions that the other trajectories do not have. One of the trajectories is randomly selected for the training dataset, while another is selected for the testing dataset. The remaining trajectory is appended to the cross-validation dataset. Furthermore, we artificially generated more training trajectories using the data points from the training dataset as described in Section \ref{proposed} Subsection D. However, we did not create artificial trajectories using the testing and cross-validation datasets. The quality of the data collected by the ESP32C3 routers is unknown, thus we use commercial off the shelf (COTS) routers as a reference to validate the quality of the data from the ESP32C3 routers. The data collected are organized into two main groups:  Dataset A is collected using the  COTS routers and Dataset B is collected using the  ESP32C3 routers.

\subsection{Dataset Information}

Dataset A contains the packets collected by COTS routers. At every position, at least 50 samples are obtained by the routers. Each sample contains a timestamp, X position, Y position, $\theta$ orientation,  Wi-Fi RSSI, Wi-Fi SQI, BLE RSSI, and BLE TX power. For the EOW 3rd floor, 11 Wi-Fi routers and 7 BLE routers are deployed to obtain the dataset. Note that some Wi-Fi routers share the same locations as the  BLE routers. The raw dataset contains approximately 500,000 samples at 1,000 different positions, where each sample has 31 wireless parameter features. For the EOW 4th floor,  9 Wi-Fi routers and 7 BLE routers are deployed to obtain the dataset. There are fewer routers on this floor due to the lack of power outlets. The raw dataset contains approximately 300,000 samples at 600 different positions, where each sample has 29 wireless parameter features. For the ECS 1st floor,  7 Wi-Fi routers and 7 BLE routers are deployed to obtain the dataset.  The raw dataset contains approximately 200,000 samples at 1000 different positions, where each sample has 24 wireless parameter features. For the ECS 5th floor,  8 Wi-Fi routers and 6 BLE routers are deployed to obtain the dataset.  The raw dataset contains approximately 300,000 samples at 600 different positions, where each sample has 24 wireless parameter features extracted from the packets.

Dataset B is sampled at the same locations and with the same procedures as Dataset A. However, Dataset B uses ESP32C3 routers, and it provides a new wireless feature known as Wi-Fi FTM. Moreover, the ESP32C3 routers occupy 40 MHz bandwidth instead of the 20 MHz bandwidth in Dataset A. These new additions increase the localization accuracy of the BiLSTM and the precision of the contact tracing algorithm.

\subsection{Identifying Unique Mobile Devices from Random MAC Addresses}

\begin{table*}[!ht]
\caption{ \label{tab:wifimacluster} Test Results of the Clustering Algorithm for Identifying Unique Mobile Devices from Random MAC Addresses. }
\begin{center}
    \begin{tabular}{|l|l|l|}
\hline
\hline
Bucket & Device & MAC Addresses\\ \hline
    0 & Galaxy S4 &  D0:22:BE:F5:7C:B4  \\ \hline
    
    1 & HTC One X &  E8:99:C4:99:57:24   \\ \hline
    
    2 & Galaxy S6 &   4E:0F:A0:57:F8:75, 26:45:19:1E:D5:FE, 1A:5B:0A:B1:7D:4A, 0E:BF:6D:4D:ED:A7, 42:B2:3B:14:49:F9, 1A:CF:16:13:A2:CB, \\ 
    & & 8C:F5:A3:3D:16:DA, 3A:DC:D3:0A:46:B6  \\ \hline

    3 & Galaxy A11 & 6A:E0:23:0C:20:0F, 56:5C:AC:D6:13:30, E2:01:19:D0:64:2D,
 FA:05:BB:EA:47:2D, A0:27:B6:EE:6A:A7, 7E:69:90:C6:C4:04, \\
 & & DA:00:FD:35:82:25, 56:2F:2B:64:BC:C5, F6:08:C4:AF:61:94,
 16:0D:FA:80:F8:1F, 5E:99:98:7B:5A:BF, 96:96:27:97:22:4C, \\
 & & FE:CB:1A:2E:F5:9A, B2:78:9D:5C:B9:1A 16:3C:FC:DF:1C:CA,
 96:38:7C:5D:20:5C   \\ \hline

    4 & iPhone SE & 82:31:01:8A:F3:AD, AE:9E:BE:7A:F3:D3, A6:E9:93:A7:9D:3E, D2:C5:A7:8B:9E:2C, 46:33:10:CE:43:3B, AA:CB:57:97:5E:5F, \\
    & & 1A:40:6D:01:B4:05, 96:C2:5B:09:D8:4E, 3A:E6:E3:9B:8E:6D, 56:D3:41:61:0B:0A, A2:68:13:44:B2:EF, 8E:6A:CF:EF:6E:1F, \\
    & & 56:A2:4A:EE:D4:46, E2:F7:83:DC:1E:E4, 22:29:5A:0D:F3:24, B6:33:3F:4F:89:1A, 9E:3D:78:F4:38:5D, ... \\ \hline

    5 & iPhone X & 86:98:6E:73:89:1D, 86:AD:C7:47:02:39, 3E:6F:2D:B3:4D:BB, 76:8A:CB:74:73:90, 9A:2D:E5:A8:F1:5A, 76:34:D2:C0:89:71, \\ 
    & & FE:B8:15:02:43:7C, 76:55:81:98:C3:78, CE:56:BC:E7:3E:72, 46:C9:78:16:41:B6, BE:F2:DB:37:1A:8A, 2A:2F:3E:B1:C7:A0, \\
    & & 6E:4C:1E:F1:8E:E8, A2:97:F2:BA:2A:D5, 6A:DD:55:59:2E:68, DA:D2:D1:55:18:60, F2:C5:62:AD:29:04, ...  \\ \hline

    6 & PinePhone & 7A:26:59:B4:C6:6D, D6:8A:05:6A:62:F5, 96:61:D9:88:25:45, 7E:53:9C:5F:BE:D0, B6:13:70:3F:28:C9, 0A:70:BB:F2:2D:9D, \\ 
    & & 52:A2:3B:BD:6D:DF, D6:DB:E0:37:8C:CE, 62:0B:F8:3C:3A:E2, BA:12:FB:78:53:F4, A2:3E:F7:DF:14:03, BE:B6:47:61:BC:31, \\
    & & EA:14:84:75:F9:00, 8E:B7:F1:4D:1A:FC, C6:41:5C:E2:C6:7B, 92:47:59:89:C4:37, CA:DC:C0:CF:39:FB, ... \\ \hline
    
    \end{tabular}
\end{center}
\end{table*}

In this section, the effectiveness of the clustering algorithm for identifying unique mobile devices from random MAC addresses is tested.  For the test setup, MAC address randomization is enabled on the devices, and they are forced to join a wireless network. Every time a mobile device joins a new wireless network, the operating system generates a new random MAC address for that specific network. Ground truth MAC addresses are obtained by looking at the settings menu. Simultaneously, the devices' probe request packets are captured at the router side. Afterwards, the clustering algorithm is applied to the probe requests to identify unique mobile devices from random MAC addresses.

Table \ref{tab:wifimacluster} shows the results of the  clustering algorithm on the testing dataset. Each row of the table contains a single bucket, of which each bucket contains MAC addresses that belong to the same mobile device.  Galaxy S4 is loaded with LineageOS 16, of which does not have MAC address randomization. Table \ref{tab:wifimacluster} shows the clustering algorithm assigning Galaxy S4's single MAC address to a single bucket and MAC addresses from other devices are not present in that bucket. The result matches the  Galaxy S4's ground truth MAC address. Similar to the Galaxy S4, the HTC One X does not have MAC randomization, and it results in a single MAC address found in  Table \ref{tab:wifimacluster}. However,  Galaxy S6 is loaded with  LineageOS 18.1, and it generates a new random MAC address upon joining a new wireless network.  Galaxy S6 is forced to join 7 different wireless networks, and the clustering algorithm places all 7 of the Galaxy S6's random MAC addresses in the same bucket. Note that the clustering algorithm has 100\% accuracy because all the MAC addresses in Galaxy S6's bucket in Table \ref{tab:wifimacluster}  matches all the MAC addresses in the ground truth. On the other hand, Android 11 on Galaxy A11 adds a new feature that randomizes MAC addresses while scanning for nearby SSIDs. The exact same test is performed on the Galaxy A11, of which the ground truth MAC addresses matches the clustering result found in Table \ref{tab:wifimacluster}. Note that the extra MAC addresses of Galaxy A11 are generated when scanning for SSIDs. We have also observed that  Android only generates a new random MAC address on the first network connection.  Rejoining a previously connected network yields the same MAC address.

The iPhone SE  supports MAC address randomization because it has iOS 15.1 firmware. For the test, iPhone SE is forced to join 7 different wireless networks, and the clustering algorithm places all 7 of the iPhone SE's random MAC addresses in the same bucket. Again, the clustering algorithm  achieves 100\% accuracy because all the ground truth MAC addresses are found at the iPhone SE's bucket in  Table \ref{tab:wifimacluster}.  The extra MAC addresses found in the iPhone SE's bucket are generated when scanning for nearby SSIDs.  The iPhone X's results are the same as the iPhone SE's results because they both have the same iOS 15.1 firmware.

The full Arch Linux distribution is installed onto the PinePhone, of which allows full control over MAC address randomization.  We wrote a script to generate 25 new random MAC addresses and to save them into a file as the ground truth.  Afterwards, the clustering algorithm's results found in  Table \ref{tab:wifimacluster} are compared to the ground truth. The clustering algorithm is able to place all the PinePhone's MAC addresses into the same bucket without any other MAC addresses from other devices being there. Overall, the clustering algorithm correctly classified every single test device into their respective buckets, even though their MAC addresses are randomized.  However, any two mobile devices that have the same model number at the same spacetime might cause the system to produce incorrect results. This is due to identical devices producing indistinguishable probe request information and RSSI information.

\subsection{Path Analysis of Dataset A}

\begin{figure*}[!t]
\centering
\includegraphics[width=\textwidth]{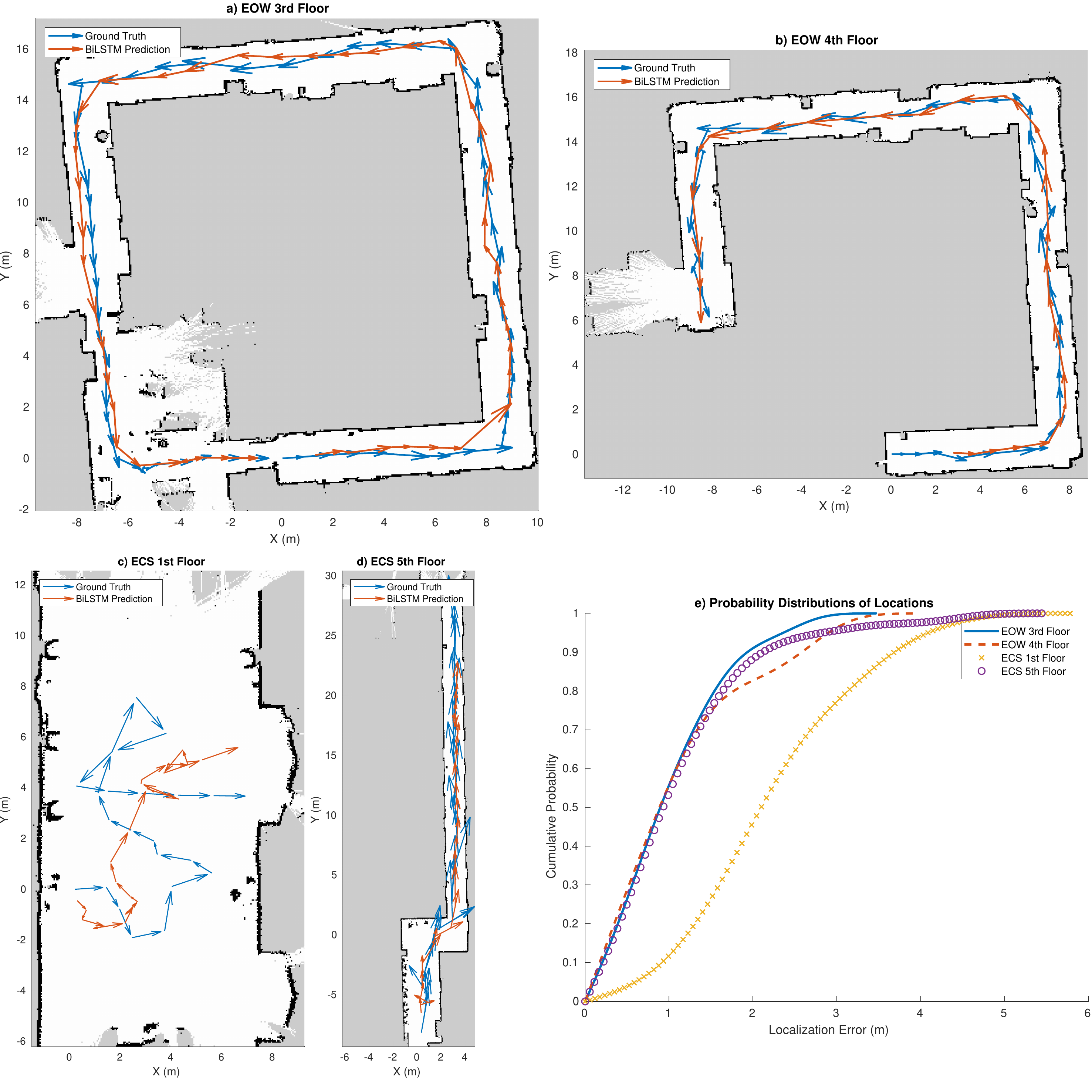}
\caption{BiLSTM's predicted locations vs ground truth at: a)  EOW 3rd floor b) EOW 4th floor c) ECS 1st floor d) ECS 5th floor e) probability distributions of Dataset A Wi-Fi+BLE } 
\label{fig:DatasetA}
\end{figure*}

\begin{table*}[!ht]
\caption{ \label{tab:BiLSTMperformance} Dataset A: BiLSTM's localization performance at different locations.}
\begin{center}
    \begin{tabular}{|l|l|l|l|l|l|l|}
\hline
\hline
Location    & Method   & APs       & RMSE (m)   & MAE (m)  &  Training Time (s)  &   Testing Time ($\mu$s)   \\ \hline
EOW 3rd Floor & BiLSTM+Wi-Fi     & 11 & 0.83 & 0.58 & 3.67 & 420 \\ \hline
EOW 3rd Floor & BiLSTM+BLE      & 7  & 0.88 & 0.56  & 3.12 & 399 \\ \hline
EOW 3rd Floor & BiLSTM+Wi-Fi+BLE & 18 & 0.82 & 0.58  & 4.34 & 445\\ \hline \hline
EOW 4th Floor & BiLSTM+Wi-Fi     & 9  & 0.92 & 0.61  & 3.55 & 410 \\ \hline
EOW 4th Floor & BiLSTM+BLE      & 7  & 0.93 & 0.63  & 3.02 & 402  \\ \hline
EOW 4th Floor & BiLSTM+Wi-Fi+BLE & 16  &  0.84 & 0.60   & 3.96 &  405 \\ \hline \hline
ECS 1st Floor & BiLSTM+Wi-Fi     &  7 & 1.69 &  1.42  & 4.72 &  340  \\ \hline
ECS 1st Floor & BiLSTM+BLE     &  7 & 2.13 & 1.73    &  3.31  & 361 \\ \hline
ECS 1st Floor & BiLSTM+Wi-Fi+BLE      & 14  & 1.30 &  1.03  & 4.32 & 417  \\ \hline \hline
ECS 5th Floor & BiLSTM+Wi-Fi     & 8  & 0.83 & 0.62   & 3.72 & 370 \\ \hline
ECS 5th Floor & BiLSTM+BLE     & 6  & 0.87 &  0.68 &  3.38 & 373 \\ \hline
ECS 5th Floor & BiLSTM+Wi-Fi+BLE     & 14  & 0.92  & 0.63 & 4.13  & 391 \\ \hline
    \end{tabular}
\end{center}
Note: Wi-Fi implies Wi-Fi RSSI and SQI,while BLE implies BLE RX power and TX power.
\end{table*}

\begin{table*}[!ht]
\caption{ \label{tab:BiLSTMperformance2} Dataset B: BiLSTM's localization performance at different locations.}
\begin{center}
    \begin{tabular}{|l|l|l|l|l|l|l|}
\hline
\hline
Location    & Method   & APs       & RMSE (m)   & MAE (m)  &  Training Time (s)  &   Testing Time ($\mu$s)   \\ \hline
EOW 3rd Floor & BiLSTM+Wi-Fi FTM    & 8 & 0.80 & 0.57 & 4.59 &  360\\ \hline
EOW 3rd Floor & BiLSTM+Wi-Fi RSSI    & 8 & 0.82 & 0.62 & 5.57 & 366 \\ \hline
EOW 3rd Floor & BiLSTM+Wi-Fi FTM+Wi-Fi RSSI & 16 & 0.75 & 0.55 & 5.93 & 442  \\ \hline \hline
EOW 5th Floor & BiLSTM+Wi-Fi FTM    & 8 & 0.75 & 0.57 & 4.21 & 461 \\ \hline
EOW 5th Floor & BiLSTM+Wi-Fi RSSI    & 8 & 0.77 & 0.59 & 3.70 & 428 \\ \hline
EOW 5th Floor & BiLSTM+Wi-Fi FTM+Wi-Fi RSSI & 16 & 0.70 & 0.52  & 3.52  &  373 \\ \hline
\hline
ECS 1st Floor & BiLSTM+Wi-Fi FTM    & 8 & 1.52 & 1.31 & 11.63 &  426\\ \hline
ECS 1st Floor & BiLSTM+Wi-Fi RSSI    & 8 & 1.63 & 1.41 & 11.89 & 409 \\ \hline
ECS 1st Floor & BiLSTM+Wi-Fi FTM+Wi-Fi RSSI & 16 & 0.89 & 0.70 & 12.07 & 446 \\ \hline \hline
    \end{tabular}
\end{center}
Note: Wi-Fi RSSI implies Wi-Fi RSSI and SQI, while Wi-Fi FTM implies RTT using IEEE 802.11mc .
\end{table*}

The BiLSTM neural networks are applied to many environments, and their RMSE and MAE performances are shown in Table~\ref{tab:BiLSTMperformance}.  At EOW 3rd floor, the  BiLSTM with Wi-Fi has a RMSE of 0.83 m and a MAE of 0.58 m. Moreover, only using BLE information yields a similar RMSE of 0.88 m and a MAE of 0.56 m because the Wi-Fi routers share the same positions as the BLE routers. Combining Wi-Fi and BLE information together results in a RMSE of 0.82 m and a MAE of 0.58 m, of which has no discernible difference. For error analysis, the ground truth trajectory is compared against the BiLSTM's predicted trajectory in Fig.~\ref{fig:DatasetA}a. The path begins at (X = 0 m, Y = 0 m) with medium error of 1.0 m.  However, the error increases to 1.5 m at the corners because that position has the least amount of LoS from the routers. Subsequently, the highest error of 2.5 m occurs in the east hallway because there are multiple objects blocking the signal paths of the routers. Afterwards, the error rapidly drops to 0.5 m as the BiLSTM recovers itself and gets back on the correct trajectory.  For the rest of the path, the error predominantly stays below 1.0 m, but there are a few locations where the error jumps above 1.0 m due to corners.

At EOW 4th floor, the  BiLSTM with Wi-Fi has a RMSE of 0.92 m and a MAE of 0.61 m. The RMSE of EOW 4th floor is slightly higher than the RMSE of EOW 3rd floor because the routers' signal paths in EOW 4th floor are blocked by more walls and doors. Moreover, the number of routers is reduced from 11 to 9 due to the lack of power outlets. Using only BLE produces a similar RMSE of 0.93 m and a MAE of 0.63 m due to the number of Wi-Fi and BLE routers being similar. Combining Wi-Fi and BLE information together results in a slightly lower RMSE of 0.84 m and a MAE of 0.60 m. The lower RMSE and MAE is caused by the increased bandwidth and the increased channel diversity. Just as before, the ground truth trajectory is compared to the BiLSTM's predicted trajectory in Fig.~\ref{fig:DatasetA}b. This time, the highest error of 2.3 m occurs at the starting position of (X = 0 m, Y = 0 m). The large error is caused by not having enough space and power outlets to place more routers at the starting position. Soon after, the error quickly decreases to  1.0 m as the BiLSTM recovers and gets back on the correct trajectory.  There are a few instances where the error jumps significantly due to the objects blocking the router's signals, but those errors are lower than the starting position errors.

ECS 1st floor is very different from the other floors because  the packets are collected in an open area instead of an enclosed hallway. In an open area, Wi-Fi RSSI and BLE RSSI  changes approximately  5 dBm per 10.0 m. The routers are not sensitive enough to detect the small changes in RSSI and creates errors in localization. Moreover, there is an extra degree of freedom compared to the hallways, of which creates more ambiguity in the trajectory. As a result, localization errors on this floor are much larger than the other floors. For Wi-Fi, the RMSE is 1.69 m and the MAE is 1.42 m, of which the localization errors are significantly higher than EOW 3rd floor, EOW 4th floor, and ECS 5th floor. Localization using BLE information has a larger RMSE of 2.13 m and a larger MAE of 1.73 m. This is caused by the BLE having a lower transmit power and worse antennas.  Combining Wi-Fi and BLE slightly lowers the RMSE to 1.30 m and the MAE to 1.03 m because the antenna diversity and the channel diversity are increased. The ground truth trajectory is compared to the BiLSTM's predicted trajectory in Fig.~\ref{fig:DatasetA}c. The trajectory starts off well at the origin of (X = 0 m, Y = 0 m) with an error of 0.5 m. However, the error rapidly increases to above 2.0 m because the predicted trajectory quickly diverges from the ground truth. The BiLSTM never recovers from incorrect predictions, and the error remains above 2.0 m. The large errors are caused by the ambiguity of the wireless features. Multiple unique positions on the map have the same Wi-Fi RSSI, SQI, and BLE RSSI.

The ECS 5th floor has a similar layout to EOW 3rd floor and EOW 4th floor because they are all located in a hallway. As a consequence, the localization performance on ECS 5th floor is very similar to the other floors. This is further evidenced by the BiLSTM with Wi-Fi on ECS 5th floor having an RMSE of 0.83 m and a MAE of 0.62 m, which is comparable to the RMSE of 0.83 m and the MAE of 0.58 m on EOW 3rd floor. Moreover, the BiLSTM with BLE on ECS 5th floor has a RMSE of 0.87 m and a MAE of 0.68 m that is similar to the RMSE of 0.88 m and the MAE of 0.56 m on EOW 3rd floor. The ground truth trajectory is compared to the BiLSTM's predicted trajectory in Fig.~\ref{fig:DatasetA}d. The predicted path has low RMSE at the origin as it is surrounded by many routers. Moreover, the localization RMSE is stable when the trajectory moves upwards. However, the RMSE jump increases drastically to 2.0 m at the end of the path due to the signal reflections at the corner of the hallway.

The cumulative distribution function (CDF) of the localization errors is shown in Fig. \ref{fig:DatasetA}e, and it tells the same story as above. The EOW 3rd floor, EOW 4th floor, and ECS 5th floor have very similar CDFs with  expected localization errors of approximately 0.89 m. Again, this is due to the floors having similar map layouts. On the other hand,  ECS 1st floor is an outlier with significantly different CDF. Most of the errors in ECS 1st floor occur between 1.0 m and 3.0 m, raising the expected localization error to 2.3 m. The significant increase in error is caused by the  ECS 1st floor having more degrees of freedom,  more possible trajectories, and ambiguity in the wireless features.

\subsection{Path Analysis of Dataset B}

\begin{figure*}[!t]
\centering
\includegraphics[width=\textwidth]{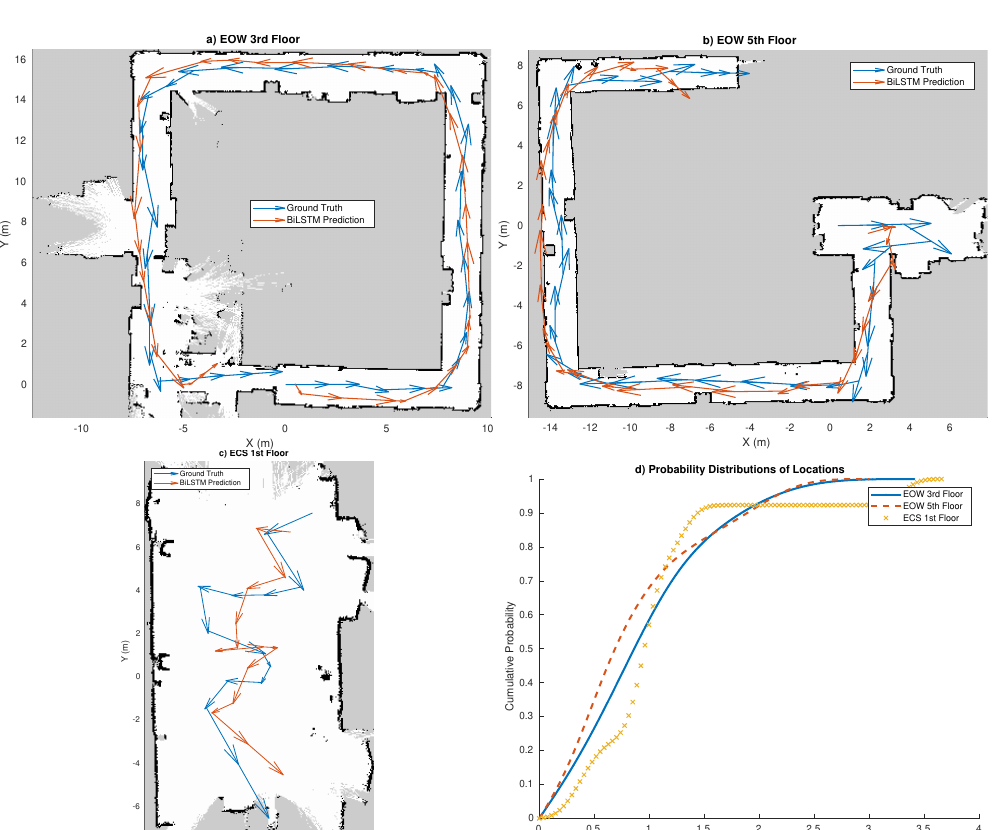}
\caption{BiLSTM's predicted locations vs ground truth at: a)  EOW 3rd floor b) EOW 4th floor c) ECS 1st floor d) probability distributions of Dataset B Wi-Fi+FTM } 
\label{fig:DatasetB}
\end{figure*}

Introducing Wi-Fi FTM to the Dataset B generally improves localization accuracy due to having more independent wireless features. For EOW 3rd floor, Table~\ref{tab:BiLSTMperformance2} shows the  BiLSTM with Wi-Fi RSSI has a RMSE of 0.82 m and a MAE of 0.62 m. This is very similar to BiLSTM and Wi-Fi RSSI in Dataset A. However, using Wi-Fi FTM yields a RMSE of 0.80 m and a MAE of 0.57 m. Combining Wi-Fi RSSI amd Wi-Fi FTM together results in a RMSE of 0.75 m and a MAE of 0.55 m, of which have slightly lower errors than Wi-Fi and BLE in Dataset A. For error analysis, the ground truth trajectory is compared against the BiLSTM's predicted trajectory in Fig.~\ref{fig:DatasetB}a. The path begins at (X = 0 m, Y = 0 m) with low error of 0.5 m. As the trajectory moves in the counterclockwise direction, the error stays below 1.0 m. However, the error increases to above 1.5 m at specific corners because the routers' LoS are broken. Moreover, the error also increases to 1.5 m at the end of the trajectory due to signal scattering in the room.

 In the open area of ECS 1st floor, the BiLSTM and Wi-Fi RSSI yields a RMSE of 1.63 m and a MAE of 1.41 m. The localization error is very large and is comparable to the BiLSTM and Wi-Fi in Dataset A.  Localization using Wi-Fi FTM has a RMSE of 1.52 m and a MAE of 1.31 m, of which has little change in error. Combining Wi-Fi RSSI and Wi-Fi FTM drastically lowers the RMSE to 0.89 m and the MAE to 0.70 m because having more independent wireless features decreases the position ambiguity in localization. Moreover, having more features increases redundancy in case one of the features is corrupted by the channel noise. The ground truth trajectory is compared to the BiLSTM's predicted trajectory in Fig.~\ref{fig:DatasetB}c. The trajectory at the origin starts a high error of 1.3 m. On the first turn, the error drops to 0.5 m. Moreover, the error hovers around 1.0 m when moving in a straight line. However, the error increases to 2.0 m at the second turn due to BiLSTM failing to predict sharp turns. Afterwards, the BiLSTM recovers and the error drops below 1.0 m for the rest of the path.

\subsection{Contact Tracing Website}

\begin{figure*}[!t]
\centering
\includegraphics[width=\textwidth]{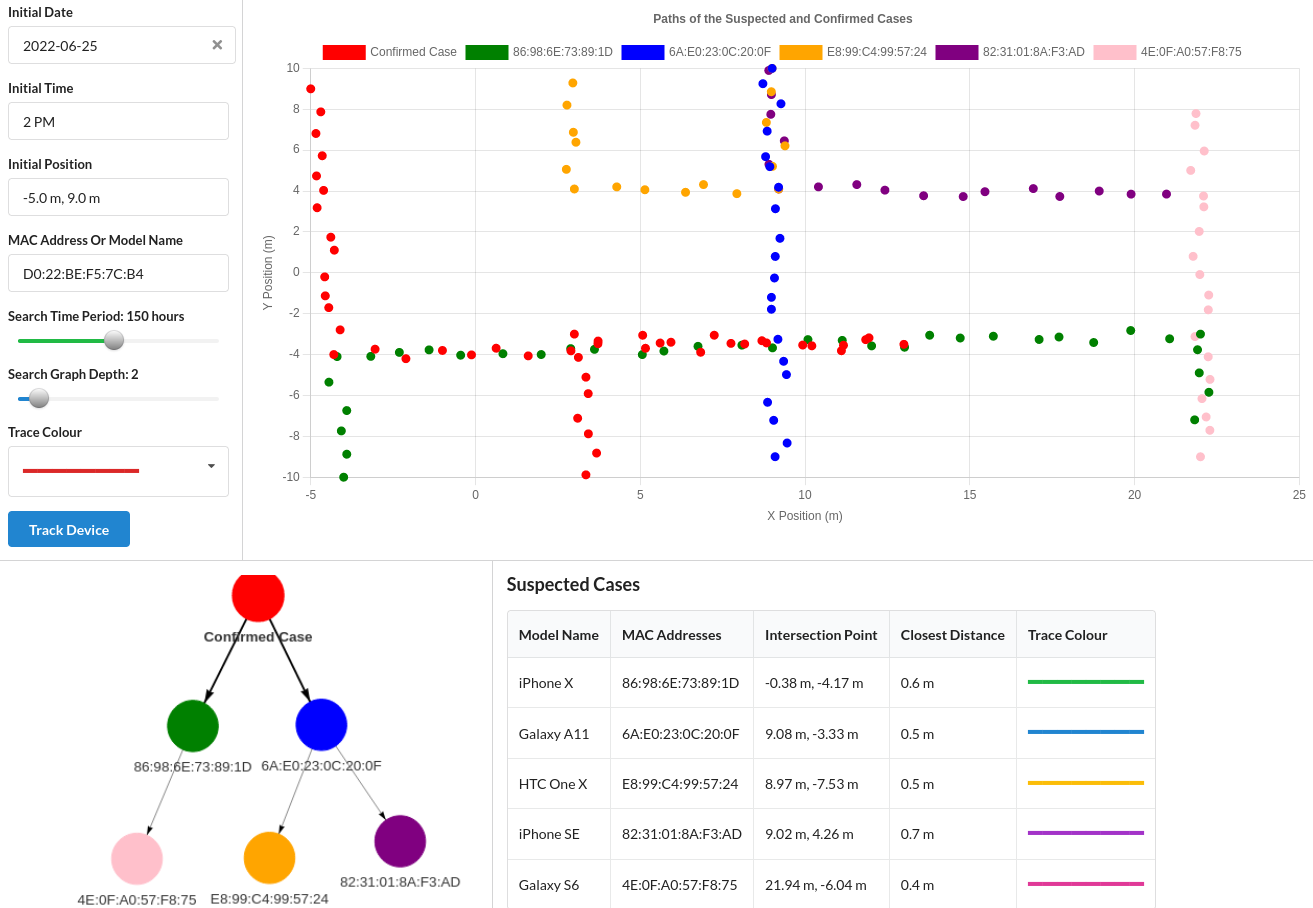}
\caption{The contact tracing website allows users to find the suspected cases using the details of the confirmed cases.} 
\label{fig:website}
\end{figure*}

We developed a contact tracing website that allows the public health authorities to find the suspected cases using the  details of the confirmed cases.  Fig.~\ref{fig:website} shows an example of the website, where the authorities enter the initial date, time, and position of the confirmed case. Afterwards, the authorities enter the MAC address/model name of the specific mobile device. The pathogen's lifetime is inputted as the search time period. A longer time period increases the search window and finds more social contacts. Subsequently, the search graph depth controls the traversal depth of the social contact graph. A graph depth of 1, lists suspected cases that have direct contact with the confirmed case. On the other hand, a graph depth of 3, lists suspected cases within 3 social contacts of the confirmed case. Moreover, the graph depth provides indirect contacts together with direct contacts.

For simplicity, the confirmed case is marked in red, while the suspected cases are marked in green, blue, orange, purple, and pink. On the top right, we plot the x and y positions of the suspected and confirmed cases.  The bottom right lists the suspected cases along with the model name, MAC addresses, intersection point, and closest distance. The red path intersects with the green path, and this is indicated in the first row of the suspected cases list, where it shows the intersection point of (-0.38 m, -4.17 m) and the closest distance of 0.6 m. Furthermore, a social contact graph is displayed in the bottom left, where it shows a link between the red circle and the green circle.

\section{Conclusion}\label{con}

We have created a novel, privacy preserving Wi-Fi and BLE contact tracing system for finding the detailed paths of the infected individuals without any user intervention. The system tracks smartphones, but it does not  require smartphone applications, connecting to the routers, or any other extraneous devices on the users. A custom built autonomous Turtlebot3 is used for site survey simulating user movement and smartphone transmission.  The smartphones' received power, transmit power, and round trip time are collected by custom ESP32C3 routers. Even though MAC randomization is employed in modern smartphones,  we have defeated it to track many devices. Afterwards, the wireless parameters collected are converted to signal path loss and ToF, of which the BiLSTM takes and predicts the absolute paths of the users. The localization performance and the RMSE is always below 0.9 m for Wi-Fi RSSI + Wi-Fi FTM. Public health authorities can use the designed website to find the paths of the confirmed cases and suspected cases, together with their MAC addresses/smartphone model specific information. They can also track indirect contact transmissions originating from surfaces and droplets.

\bibliography{citelist}

\EOD

\end{document}